# Characteristics of Al/Ge Schottky and Ohmic contacts at low temperatures

**Shreyas Pitale,**[a] **Manoranjan Ghosh,**[a]***** **S.G. Singh,**[a] **Husain Manasawala,**[a] **G.D. Patra,**[a] **Shashwati Sen**[a,b]

[a]*Crystal Technology Section, Technical Physics Division, Bhabha Atomic Research Centre, Mumbai-400085, India*

[b]*Homi Bhabha National Institute, Mumbai-400094, India*

*Email: mghosh@barc.gov.in

**Abstract:**

Schottky barrier contact has been fabricated by thermal deposition of Al on (100) Ge (impurity concentration~$10^{10}$/cm$^3$ at 80K) that shows extrinsic p-type to intrinsic n-type transition near 180K. Both p and n-type Ge exhibits ideal Schottky behaviour with low reverse current and near unity ideality factors obtained from the linear form of temperature dependent current-voltage (I-V) characteristics. The diode current at various temperatures change its direction at non-zero applied bias that reflects a shift in position of charge neutrality level (CNL) from the Fermi level of Ge. With the rise in temperature, Schottky barrier height (SBH) steadily increases for p-Ge that can be understood on the basis of observed variation in CNL. Values of SBH determined from the zero bias Richardson plot agrees well with that estimated from the Schottky-Mott rule for strongly pinned interface. Activation energies are determined from the Richardson plot at various forward voltages and found to decrease with applied bias for n-Ge but reduces to zero for p-Ge that shows work function similar to Al. Annealing of Al/Ge induces regrowth of p-type Al doped Ge layer that exhibits gradual reduction of Al concentration towards p-Ge crystal. Al doped Ge(P$^+$)/Ge (P) junction thus fabricated shows linear current-voltage (I-V) characteristics in the extrinsic region (below 180K). In the intrinsic region (above 180K), rectification is observed in the I-V curve due to temperature dependent change in conductivity of both Al doped Ge layer and Ge crystal.

## 1. Introduction:

Fabrication of electrical contacts on semiconductor is required for its application in electronic devices. Due to the presence of a potential barrier between metal and semiconductor, a Schottky barrier is created that prevents flow of charge carriers thereby forming a diode with high degree of rectification and faster transitions. Metal/Ge Schottky contacts were widely investigated [1-4] and barrier height is found to be independent of metal work function due to Fermi level pinning near charge neutrality level [5,6]. Al is a popular choice for making Schottky contacts on Ge and reported to exhibit higher than predicted values of barrier heights and ideality factor [1,7-9]. Impurity concentration of germanium employed for making Al/Ge Schottky contacts were greater than $10^{14}/cm^3$ and high quality diode showing near unity ideality factor along with barrier height close to the difference between electron affinity of Ge and Al work function has not been achieved. Fabrication of ideal metal/germanium Schottky contact remained as challenge due to dependence on various parameters such as interface, surface states and band structure modification [10-12]. Further, electrical conductivity of nominally doped Ge is a function of temperature [13] that has influence on the temperature dependence of metal/Ge Schottky barrier parameters. Characteristics of Al/Ge Schottky contact at low temperatures have not been studied in detail.

Reduction of Shottky barrier height (SBH) and fabrication of Ohmic contacts is an important step to get signals into and out of a semiconductor. SBH of Metal/Ge can be reduced by inserting a dielectric layer between metal and germanium [14-16]. Barrier height values such as 0.34 eV, 0.15 eV and 0.35 eV are achieved for Ti germanide [17], sulfur-passivated Ni germanide [18] and Fe/MgO contacts [19] respectively and found to be as low as 0.13eV and 0.09 eV for Al contacts on Ge by $CF_4$ plasma treatment [8] and introduction of few monolayers of crystalline and amorphous $Ge_3N_4$ [5]. It is claimed that Fermi level pinning at the interface between *n*-Ge and a metal leads to the formation of a Schottky barrier but for *p*-type Ge it works opposite and all metal contacts show Ohmic behaviour [7]. Further, Al shows good Ohmic contacts on n-type Ge and hole barrier height of 0.57eV to p-type Ge after sulphur passivation treatment [20]. A transition from rectifying to Ohmic behavior is observed in Al/n-type Ge contact with $O_2$ plasma treatment [21]. Investigation on Al Schottky barrier on pristine p and n-type Ge is relevant for understanding large variation in heights observed for doped Ge as well as after interface modification.

Barrier height of metal/semiconductor Schottky contact can be reduced and transformed to Ohmic enough by annealing that creates an alloy between the semiconductor and the metal at the junction [22,23]. Post-metallization annealing in $N_2$ ambience of Al/p-Ge(100) junctions is found to be an effective means of controlling the Schottky barrier height [24]. A reduced barrier height around 0.37 eV is obtained at the epitaxial-NiGe$_2$/Ge(100) interface formed by pulsed laser annealing and attributed to lowering of density of interface states due to reduced number of dangling bonds [25]. A heavily doped (conc. > $10^{19}$/cm$^3$) interfacial layer creates a thin depletion region that that offer very low barrier height due to tunneling of carriers [26]. A low resistivity contact has been achieved by the growth of heavily doped semiconductor thin layer through germanidation at the Ni/Ge interfaces [27]. In a separate study, Ni/Ge and PdGe/Ge Schottky and Ohmic contacts have been fabricated by germanide formation through annealing [28]. These contacts were also observed to remain stable over a wide temperature range during annealing. Al has been widely used for making Schotky contacts on Ge as well as regrowth of epitaxial Al doped Ge interfacial layer by thermal annealing [29]. However formation of Ohmic contacts on Ge with Al doped Ge interfacial layer is not reported earlier and temperature dependent characteristics of electrical contact formed by the combination of Ge and Al doped Ge remain unexplored.

The current study aims to address few important aspects of Al contact on germanium crystal. Pristine as well as doped Ge exhibit change in type of conductivity due to variation in temperature that determines the contact properties. Therefore, study of contact properties at low temperature is important particularly when metal contact is taken on high pure germanium and heavily doped Ge interfacial layer is used to make Ohmic contact. Here, Al is deposited on high pure germanium substrate (impurity conc. ~$10^{10}$/cm$^3$) to make ideal Schottky barrier (ideality factor~1 and barrier height 0.15 eV) diode which was hitherto not been achieved. Experimentally determined barrier heights matches with the theoretically predicted values and its temperature dependence is explained by the observed variation in charge neutrality level given by voltage required to reverse the direction of the diode current. Sudden change in temperature dependence of barrier parameters is observed when Ge undergoes transition from extrinsic p-type to intrinsic n-type conductivity. Further, Al doped Ge (p+) layer is grown on to p-type Ge by annealing Al/Ge Schottky contacts at 350$^0$C followed by slow cooling. Regrown Ge in top Al layer creates heavily doped (p$^+$) thin Al-Ge layer at the interface. In the extrinsic region (below 180K), both Al doped Ge and Ge substrate are found to be p-type by Hall measurement and the resulting p+/p junction shows Ohmic behaviour. At higher temperature, current-voltage characteristics exhibit varying degree of rectification depending on the conductivity of Al doped Ge and Ge substrate underneath.

## 2. Experimental

Commercially available Ge (100) crystal (acceptor conc. ~$10^{10}$/cc at 77K) was cut by diamond wheel to pieces having flat faces and desired size of 8mm × 8mm × 1.5 mm and lapped by SiC abrasives sheets successively with grit sizes from 220 to 1500. Lapped surfaces were ultra-sonicated in methanol for 15 minutes followed by polish etching in 3:1 solution of $HNO_3$:HF (by volume) for 3 minutes. For making Schottky contacts, approximately 80-90 nm thick Al films was deposited on one (100) face of Ge under chamber pressure down to $2\times10^{-6}$ mbar. For the regrowth of Al doped Ge layer, in-situ annealing of as deposited Al/Ge was performed immediately after thermal evaporation in an argon atmosphere at $350^0$C for 40 minutes followed by slow cooling to room temperature during 90 minutes. The easily removable Al layer converts in to rugged p-type Al-Ge contacts on Ge crystal.

Thickness of Al films were determined by Taylor-Hobson make optical 3D profiler (Model-CCI-MP) based on coherence scanning interferometry. Hall coefficient of Al-Ge samples were measured by Ecopia HMS5000 Hall effect measurement system within a range of temperatures from 77K to 300K. Four contacts were taken on the corners of square sample through In-Ga eutectic in Van der Pauw configuration. The sheet resistance of the samples was measured by collinear 4-probe method.

The device having material combination Al/Ge/In-Ga eutectic (Fig. 1) was mounted in a cryostat for temperature dependent current-voltage (I-V) measurement by a Keithley 6517B electrometer. In-Ga eutectic was used to make Ohmic back contacts on Ge. Cryostat was evacuated and cooled to 77K using liquid nitrogen and controlled using a PID temperature controller. Analysis of results is carried out in line with the band theory of metal/semiconductor junction and Schottky diode current equation based on thermo-ionic emission [30].

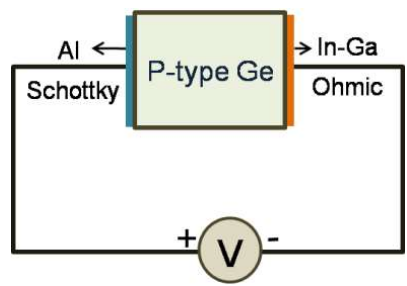

**Fig. 1. Configuration of Al/p-Ge/In-Ga Schottky diode.**

## 3. Results and Discussion

*3.1 Analysis of current-voltage (I-V) characteristics of Al/Ge contacts by Schottky barrier model*

I-V characteristics of un-annealed Al/Ge/In-Ga device (Fig. 2) in the temperature range 80-300 K shows excellent rectification that indicates formation of high quality metal semiconductor Schottky barrier diode. As the temperature increases, both forward and reverse current attains higher values and knee voltages shift to lower values due to thermally generated carriers. The high pure germanium used here exhibits p-type conductivity in the extrinsic region below 180K and exhibits low reverse current along with exponentially rising forward current, resulting rectification ratio (RR) as high as $10^4$. In the intrinsic region (above 180K), Ge shows n-type conductivity and thermally generated minority carriers contribute to higher reverse current that reduces the RR although Schottky contact is formed by Al on both p and n-type Ge.

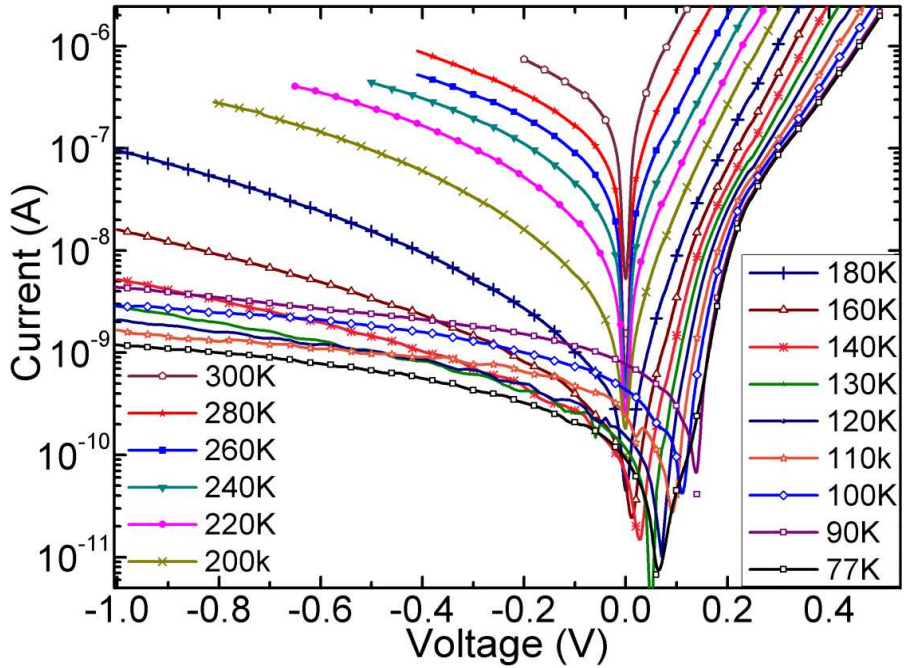

**Fig. 2. Temperature dependent semi-logarithmic I-V characteristics of Al/Ge/In-Ga Schottky diode.**

Temperature dependent Schottky barrier parameters are determined by analyzing I-V characteristics in accordance with the developed theory of Schottky barrier diodes [31,32]. The current through the Schottky diode (I) can be expressed by Shockley equation based on thermionic emission,

$$I = I_o e^{\frac{qV}{nkT}}\left(1 - e^{\frac{-qV}{kT}}\right) \quad (1)$$

where V is the applied bias across the junction, q is the electronic charge (1.6×10⁻¹⁹ C), k is the Boltzmann's constant (1.38×10⁻²³ J/K) and T is the absolute temperature in Kelvin and $I_o$ is the reverse saturation current ensuing from the straight line intercept of the *ln*(I) vs. V plot at V = 0. $I_o$ can be expressed by the Richardson-Dushman equation as:

$$I_o = AA^*T^2 e^{\frac{-q\phi_b}{kT}} \quad (2)$$

where, A is the contact surface area, A* is the Richardson constant, $\phi_b$ is the zero bias effective barrier height and *n* is the ideality factor or quality factor which is a measure of conformity with ideal diode behavior. *n* can be experimentally determined from the slope of the ln(I) vs. V characteristics given by

$$n = \frac{q}{kT}\left(\frac{dV}{d(\ln I)}\right) \quad (3)$$

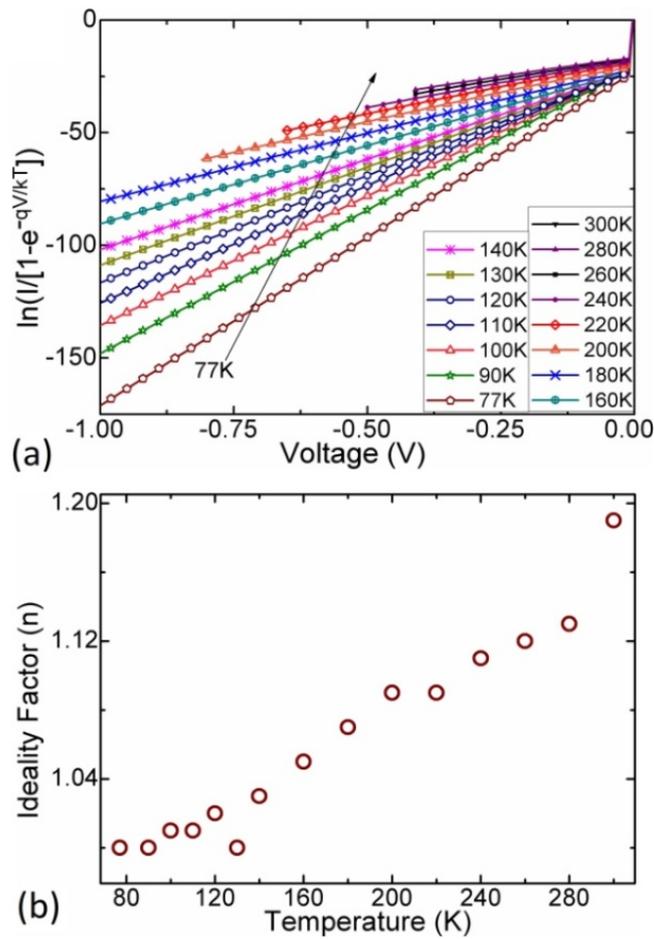

**Fig. 3.** (a) ln(I) vs. V plot of Al/Ge Schottky contact at various temperatures. (b) Temperature dependent ideality factor (n).

The equation (1) can be converted to a linear form

$$ln\left(\frac{I}{\left[1-e^{\frac{-qV}{kT}}\right]}\right) = ln(I_o) + \frac{qV}{nkT} \qquad (4)$$

Eq. 4 is used to fit the obtained temperature dependent I-V data of the Al/Ge contact as shown in Fig. 3(a) that shows good linear characteristics in the reverse biased region. The inverse of the slopes of these lines, $dV/d[ln\{I/(1-e^{-qV/kT})\}]$ were used to calculate the ideality factor ($n$) using Eq. 3. Values of n are found to be sufficiently close to ideal values (1.0 to 1.1) as conformity to the ideal diode behavior. Ideality factor steadily increases with the increase in temperature due to rise in reverse saturation current contributed by thermally generated minority carriers [Fig. 3(b)].

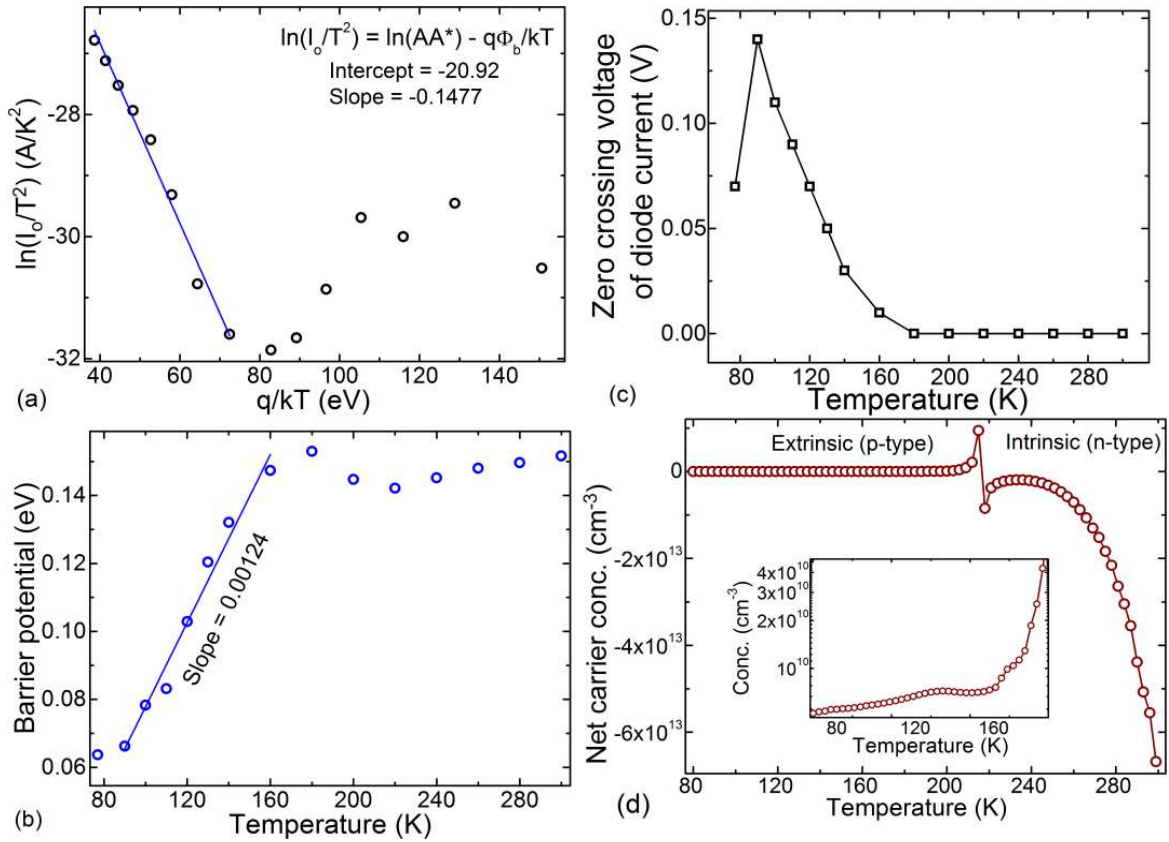

**Fig. 4. (a) Richardson plot at zero bias with reverse saturation current ($I_0$) calculated from the linear form of diode current, (b) Temperature dependent Schottky barrier potential $\Phi_b$ (eV), (c) Zero current voltages at various temperatures in the current-voltage characteristics, (d) Net carrier concentration of pristine Ge crystal in the temperature range 77-300K.**

The reverse saturation current ($I_o$) can be calculated from the y-intercept of straight line plot obtained using Eq. 4 for all measured temperatures. Taking natural logarithm of Richardson-Dushman equation 2, values of $I_o$ can be further used to draw a zero bias Richardson plot [$ln(I_o/T^2)$ vs q/kT] as shown in Fig. 4(a). Richardson plot is linear in the intrinsic temperature range of Ge under consideration (180-300K) where electrical conduction is dominated by thermally generated carriers. In the extrinsic region (80K –180K, q/kT > 80) thermal generation of carriers is unlikely and the plot shows deviation from the linear dependence as the impurity driven electrical conduction is majorly independent of temperature. The y-intercept of the linear part of the Richardson plot $ln(AA^*)$ is determined experimentally as shown in Fig. 4(a). The value of $lnAA^*$ was found to be -20.92 and the Richardson-Dushman equation is directly used to calculate the temperature dependent barrier height, $\Phi_b$ with known values of $AA^*$ and $I_o(T)$.

As shown in Fig. 4(b), $\Phi_b$ increases linearly with temperature (co-efficient 0.00124 eV/K) in the extrinsic p-region followed by stable higher values (0.14 eV - 0.16 eV) in the intrinsic n-region. Metal/Ge Schottky contact is reported to exhibit strong Fermi level pinning effect around the charge neutrality level (CNL) of Ge. In that case Schottky barrier height can be expressed by $\phi_b = S(\phi_m - \phi_{CNL}) + (\phi_{CNL} - \chi_S)$ where $S$ is $d\phi_b/d\phi_m$ is the pinning factor and $\chi_S$ and $\phi_{CNL}$ are the semiconductor electron affinity and the charge neutrality level, respectively, measured from the vacuum level. In case of maximum pinning (S=0), $\phi_b = (\phi_{CNL} - \chi_S)$ is determined mainly by the position of the CNL. The CNL of surface states is the position for the Fermi level which renders the surface without a net charge. When the Fermi level is above the CNL, the surface is negatively charged and vice versa. For a charge neutral Ge surface, current through Al/Ge should be zero at zero applied voltage. Any shift from the zero voltage to reverse the current direction reflects deviation from the surface charge neutrality. The position of CNL can be experimentally found out from the I-V curves showing negative currents at finite positive voltages and changes its direction with non-zero applied bias in case of p-type Ge (Fig. 2). With the increase in temperature the required voltage for current reversal gradually decreases and becomes zero for n-type Ge as plotted in Fig. 4(c). Thus CNL matches with the Fermi level for n-type Ge and $\phi_{CNL} \approx \phi_S$ [Fig. 5 (b)]. For p-type Ge, surface is positively charged, Fermi level is below CNL and $\phi_{CNL}$ as measured from the vacuum level is lower than $\phi_S$ [Figure 5(d)]. As shown in Fig. 4(c),

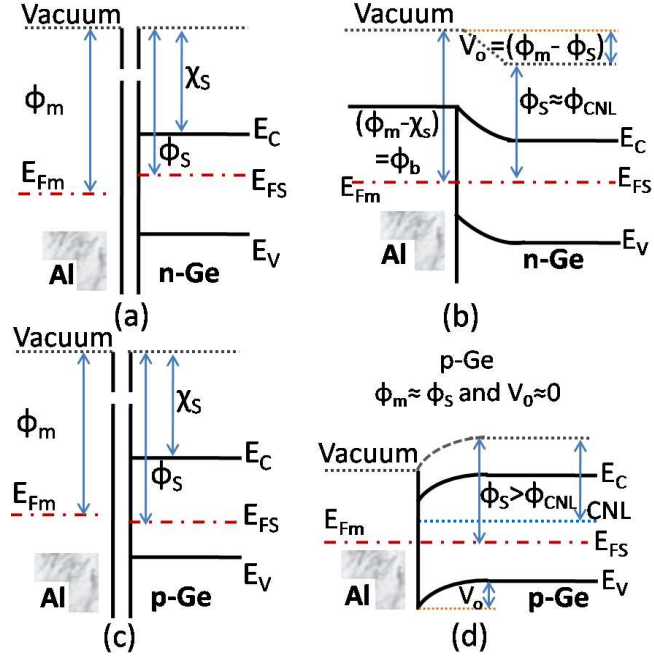

**Figure 5:** Band diagram of Schottky barrier before [(a), (c)] and after [(b), (d)] the contact for n and p-type Ge.

CNL reduces ($\phi_{CNL}$ increases) and approaches towards Fermi level with the increase of temperatures in p-type Ge. Thus, for a fixed value of $\chi_S$, temperature dependence of $\phi_b = (\phi_{CNL} - \chi_S)$ [S=0] follows $\phi_{CNL}$ which is opposite to that observed for CNL. It should be mentioned that CNL holds a linear relation with the impurity concentration of a semiconductor [33]. As shown in Fig. 4 (d), the impurity concentration of Ge used here gradually increases with temperature from $5 \times 10^9/cm^3$ (at 80K) to $\sim 10^{14}/cm^3$ (at 300K) showing p-n transition near 180K. The variation of $\phi_b$ with temperature follows similar trend of net impurity concentration. For n-type Ge, $\phi_b$ attains higher value and slowly increases with temperature as there is minor variation in $\phi_{CNL}$ and the rise in net carrier concentration is mainly due to thermal generation from Ge lattice. The values of $\phi_b$ obtained from Richardson plot also matches with that calculated from the experimentally observed CNL and standard values of $\phi_S \approx \phi_m$ (4.2 eV) and $\chi_S$ (4.0 eV). For CNL values $\sim 0.15$ eV (at 90K), $\phi_b = (\phi_m - CNL - \chi_S)$ [S=0] = 0.05 eV is similar to that determined from the Richardson plot ($\sim 0.06$ eV). In case of n-Ge, CNL $\approx 0$ eV and $\phi_b$ attains a stable value around 0.2 eV. Barrier height and ideality factors close to the theoretically predicted values are obtained mainly due to the purest form of the Ge ($\sim 10^{10}/cm^3$) employed in this investigation and rarely reported in the literature [8].

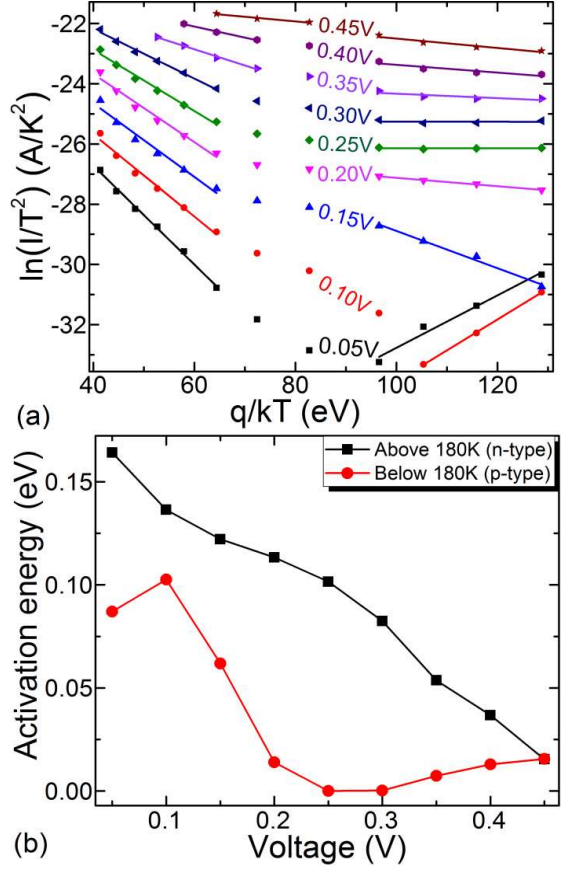

**Fig. 6. (a) Richardson plots at various forward bias voltages. (b) Activation energy as determined from the slope of the voltage dependent Richardson plot.**

Voltage dependent activation energy for both extrinsic (p-type) and intrinsic (n-type) regions can be determined from the Richardson plot at various forward bias voltages [31]. Richardson-Dushman equation for any fixed bias voltage can be written as:

$$ln\left(\frac{I(V)}{T^2}\right) = ln(AA^*) - \frac{q\Phi_b(V)}{kT} \qquad (5)$$

The Ln(I/T$^2$) vs q/kT plot for various forward bias voltages show two distinct linear regions [Fig. 6(a)]. Activation energies ($V_0$) determined from the slope of the voltage dependent Richardson plot are shown in Fig. 6(b). For q/kT <70 (intrinsic n-type) $V_0$ varies from 80 meV to 160 meV and decreases with the increase in applied biases in line with the previous report [34]. In case of q/kT >70 (extrinsic p-type) low values of activation energies are obtained except for forward voltages lower than CNL that give rise to negative current. Nearly zero values of $V_0$ observed in p-type Ge indicates $\phi_S \approx \phi_m$ [Fig. 5(d)] and negligible band bending.

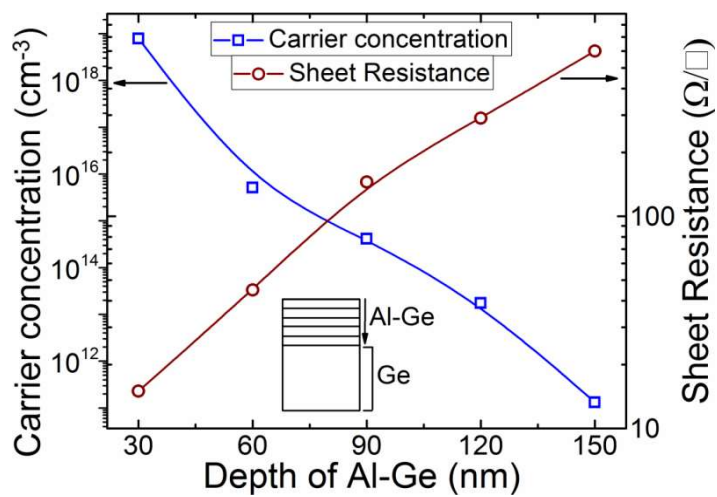

**Fig. 7.** Variation of Al impurity concentration and sheet resistance along the depth of Al-Ge regrown layer from its top surface.

*3.2 Characteristics of Al-Ge regrown contact on Ge*

Heat treatment of metal/germanium contact during or after deposition significantly changes the barrier height and current voltage characteristics [18,35]. Germanium is likely to form alloy with the metal that creates an interfacial layer. One more possibility is exchange of atoms and regrowth of metal doped epitaxial layer at the interface. In our previous work, a process for fabrication of heavily doped p type contact based on solid-state re-growth of Al on Ge is established [36]. Al/Ge Schottky contact is annealed at 350°C followed by slow cooling to regrow Al doped p-type Ge on to the p-type Ge crystal. Net carrier concentration and sheet resistance along the depth of the regrown layer has been measured and plotted in Fig. 7. Al impurity concentration (sheet resistance) decreases (increases) exponentially from the top surface and essentially form Al-Ge (highly doped)/Ge junction showing continuous variation in Al impurity concentration. The I-V characteristic of such type of contact shows broadly non-rectifying characteristics ruling out the presence of abrupt metal/semiconductor interface (Fig. 8). Only within a narrow range of temperature (180-260K), the I-V curves show dissimilar current in the forward and reverse characteristic indicating a diode like behavior. The variation in I-V characteristics with temperature can be understood in terms of the type of conductivity of the regrown Al-Ge layer and the Ge crystal underneath at a particular temperature.

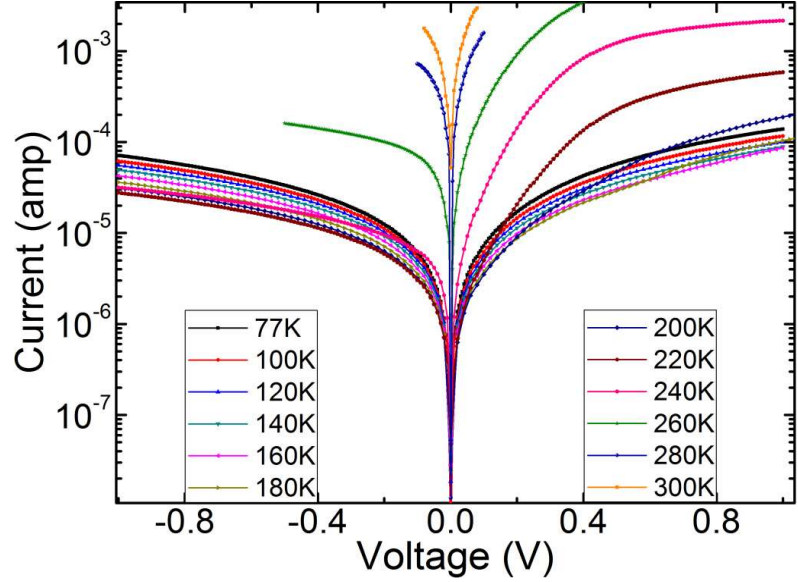

**Fig. 8.** Semi-logarithmic I-V plot at various temperature of Al-Ge/Ge contact by obtained by thermally induced regrowth of Al/Ge at 350°C.

Temperature dependent Hall co-efficients have been measured along the depth of Al doped Ge regrown layer. Surfaces at different depth from the top surface have been exposed by successive chemical mechanical polishing [36]. Nominally doped Ge is found to exhibit extrinsic p-type to intrinsic n-type transition at certain temperature depending on the impurity concentration of the crystal [13]. For undoped Ge, the transition from p to n type conductivity occurs around 180 K [Fig. 9(a)]. The Al doped Ge also exhibits similar type transition in conductivity at higher temperatures depending on Al impurity concentration. For a measured Al concentration ($\sim 10^{11}/cm^3$) at the depth of around 150 nm (bottom layer) from the top surface, the regrown layer shows p to n type transition at 220K. Therefore, three different combinations of conductivities in Al doped Ge/Ge junction is created within the temperature range from 80 to 300K as shown in Fig. 9(a). Below 180K, both Al doped Ge and Ge crystal are p-type and p+/p type junction is formed. Above 250K, both Ge and Al doped Ge are n-type in nature and formation of n+/n junction can be seen. Within the temperature range 180 to 250 K, Ge crystal is n-type but Al doped Ge is p-type resulting p+/n type junction. The nature of I-V characteristics clearly follows the change in properties of the junction due to difference in conductivity of material combination at various temperatures. The rectification ratio (ratio of forward and reverse current) is calculated from the I-V curves at various temperatures and plotted in Fig. 9(b) that shows distinct behaviors for three different types of junctions described above. No rectification is seen in the range of 80 to 180K, the p+/p region where the I-V characteristics are mostly Ohmic in nature. In the temperature range 180-

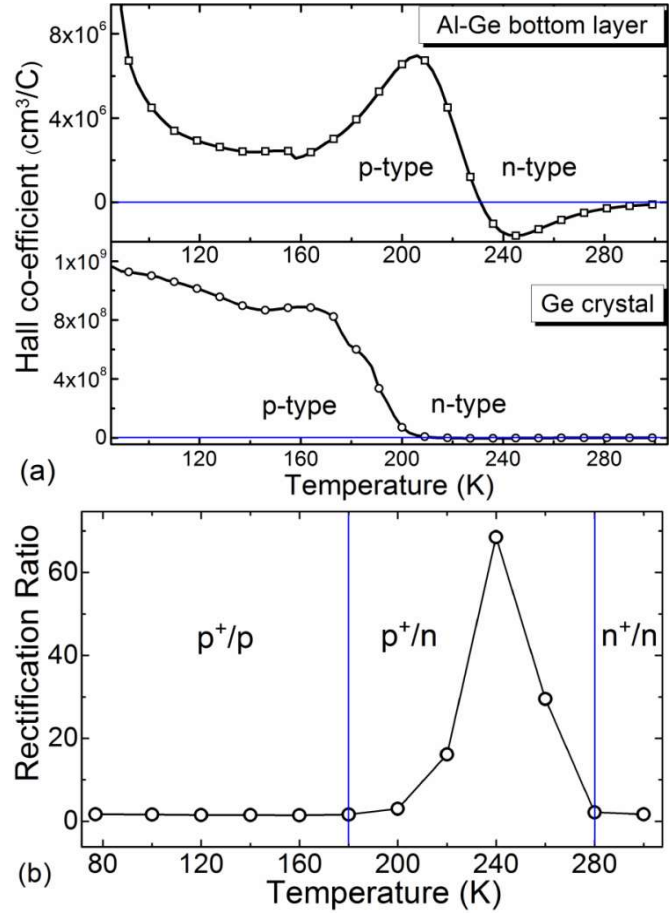

**Fig. 9.** (a) Temperature dependent Hall co-efficient at the bottom of the regrown Al-Ge layer and pristine Ge (100) crystal. (b) Rectification ratio of Al-Ge/Ge contact calculated from the forward and reverse current at 0.1V as a function of temperature.

260K, I-V curves exhibit diode behavior with high degree of rectification due to the formation of p-n junction. Rectification ratio again reduces and linear I-V characteristics are observed in the temperature range 260-300K where n+/n type of junction has been formed. It should be mentioned that no barrier is formed in case of p+/p or n+/n junctions and carrier transport is primarily diffusion controlled from higher to lower concentration.

## 4. Conclusions

Al/Ge Schottky barrier diode has been fabricated and temperature dependent barrier parameters are analyzed from the I-V characteristics. High purity germanium used in this study produces ideal Schottky contact with Al showing barrier potential as well as ideality factor close to the theoretically predicted values. Temperature dependence of Al/Ge Schottky barrier height has been explained on the basis of charge neutrality level of Ge in both extrinsic (p-type) and intrinsic (n-type) regions. Shottky diode performance cease to exist for heat treated Al/Ge contact due to regrowth of Al doped Ge layer at the interface that shows smooth variation of Al impurity concentration along the depth. I-V characteristics of such junction exhibits linear behavior in the temperature range where both Al doped Ge and Ge crystal underneath have similar type of conductivities. Within a narrow range of temperature, Al doped Ge is p-type but Ge crystal shows n-type conductivity and the junction exhibits diode type of behavior with high degree of rectification.

**Acknowledgment**

Authors are thankful to Dr. T. V. C. Rao for his constant encouragement and support. Unconditional help received from all members of Crystal Technology Section is gratefully acknowledged.

**References**

[1] A. Thanailakis, D. C. Northrop, Metal-germanium Schottky barriers, Solid State Electronics 16 (1973)1383-1389, https://doi.org/10.1016/0038-1101(73)90052-X

[2] J. Wang, W. Huang, J. Xu, J. Li, S. Huang, C. Li, S. Chen, Schottky barrier height modulation effect on n-Ge with TaN contact, Materials Science in Semiconductor Processing, 91 (2019) 206-211, https://doi.org/10.1016/j.mssp.2018.11.016

[3] M. Kobayashi, A. Kinoshita, K. Saraswat, H.S. Philip Wong, Y. Nishi, Fermi level depinning in metal/Ge Schottky junction for metal source/drain Ge metal-oxide-semiconductor field-effect-transistor application, J. Appl. Phys. 105 (2009) 023702, https://doi.org/10.1063/1.3065990

[4] A. Chawanda, K.T. Roro, F.D. Auret, W. Mtangi, C. Nyamhere, J. Nel, L. Leach, Determination of the laterally homogeneous barrier height of palladium Schottky barrier diodes on n-Ge (1 1 1), Materials Science in Semiconductor Processing 13 (2010) 371-375, https://doi.org/10.1016/j.mssp.2011.05.001

[5] R. R. Lieten, V. V. Afanas'ev, N. H. Thoan, S. Degroote, W. Walukiewicz, and G. Borghs, "Mechanisms of Schottky barrier control on n-type germanium using $Ge_3N_4$ interlayers, Journal of The Electrochemical Society, 158 (4) H358-H362 (2011), https://iopscience.iop.org/article/10.1149/1.3545703

[6] A. Dimoulas, P. Tsipas, A. Sotiropoulos, E. K. Evangelou, Fermi-level pinning and charge neutrality level in germanium, Appl. Phys. Lett. 89 (2006) 252110, https://doi.org/10.1063/1.2410241

[7] R. R. Lieten, S. Degroote, M. Kuijk, G. Borghs, Ohmic contact formation on n-type Ge, Appl. Phys. Lett. 92 (2008) 022106, https://doi.org/10.1063/1.2831918

[8] J.-R. Wu, Y.-H. Wu, C.-Y. Hou, M.-L. Wu, C.-C. Lin, L.-L. Chen, Impact of fluorine treatment on Fermi level depinning for metal/germanium Schottky junctions, Appl. Phys. Lett. 99 (2011) 253504, https://doi.org/10.1063/1.3666779

[9] D. R. Gajula, P. Baine, M. Modreanu, P. K. Hurley, B. M. Armstrong, D. W. McNeill, Fermi level de-pinning of aluminium contacts to *n*-type germanium using thin atomic layer deposited layers, Appl. Phys. Lett. 104 (2014) 012102, https://doi.org/10.1063/1.4858961

[10] A. Toriumi, T. Tabata, C. H. Lee, T. Nishimura, K. Kita, K. Nagashio, Opportunities and challenges for Ge CMOS – Control of interfacing field on Ge is a key, Microelectronic Engineering, 86 (2009) 1571-1576, https://doi.org/10.1016/j.mee.2009.03.052

[11] A. Dimoulas, P. Tsipas, Germanium surface and interfaces, Microelectronic Engineering, 86 (2009) 1577-1581, https://doi.org/10.1016/j.mee.2009.03.055

[12] I. Jyothi, V. Janardhanam, V. R. Reddy, C.-J. Choi, Modified electrical characteristics of Pt/n-type Ge Schottky diode with a pyronine-B interlayer, Superlattices and Microstructures 75 (2014) 806-817, https://doi.org/10.1016/j.spmi.2014.09.016

[13] M. Ghosh, S. Pitale, S.G. Singh, S. Sen, S.C. Gadkari, Impurity concentration dependent electrical conduction in germanium crystal at low temperatures, Bull. Mater. Sci. 42 (2019) 264, https://doi.org/10.1007/s12034-019-1944-8

[14] Y. Zhou, M. Ogawa, X. H. Han, K. L. Wang, Alleviation of Fermi-level pinning effect on metal/germanium interface by insertion of an ultrathin aluminum oxide, Appl. Phys. Lett. 93 (2008) 202105, https://doi.org/10.1063/1.3028343

[15] T. Nishimura, K. Kita, A. Toriumi, A Significant shift of Schottky barrier heights at strongly pinned metal/germanium interface by inserting an ultra-thin insulating film, Appl. Phys. Exp.1(5) (2008) 051406, https://iopscience.iop.org/article/10.1143/APEX.1.051406

[16] G. Liu, M. Zhang, Z. Xue, X. Hu, T. Wang, X. Han, Z. Di, Fermi level depinning in Ti/n-type Ge Schottky junction by the insertion of fluorinated grapheme, Journal of Alloys and Compounds, 794 (2019) 218-222, https://doi.org/10.1016/j.jallcom.2019.04.174

[17] D. D. Han, Y. Wang, D. Tian, W. Wang, X. Y. Liu, J. F. Kang, R. Q. Han, Studies of Ti- and Ni-germanide Schottky contacts on *n*-Ge(1 0 0) substrates, Microelectronic Engineering 82 (2005) 93-98, https://doi.org/10.1016/j.mee.2005.06.004

[18] K. Ikeda, Y. Yamashita, N. Sugiyama, N. Taoka, S. Takagi, Modulation of NiGe⁄GeNiGe⁄Ge Schottky barrier height by sulfur segregation during Ni germanidation, Appl. Phys. Lett., 88 (2006) 152115 https://doi.org/10.1063/1.2191829

[19] Y. Zhou, W. Han, Y. Wang, F. Xiu, J. Zou, R. K. Kawakami, K. L. Wang, Investigating the origin of Fermi level pinning in Ge Schottky junctions using epitaxially grown ultrathin MgO films, Appl. Phys. Lett. 96 (2010) 102103, https://doi.org/10.1063/1.3357423

[20] A.V. Thathachary, K. N. Bhat, N. Bhat, M.S. Hegde, Fermi level depinning at the germanium Schottky interface through sulphur passivation, Appl. Phys. Lett. 96 (2010) 152108, https://doi.org/10.1063/1.3387760

[21] V. Janardhanam, H.-J. Yun, I. Jyothi, S.-H. Yuk, S.-N. Lee, J. Won, C.-J. Choi, Fermi-level depinning in metal/Ge interface using oxygen plasma treatment, Applied Surface Science 463 (2019) 91–95, https://doi.org/10.1016/j.apsusc.2018.08.187

[22] A. Chawanda, C. Nyamhere, F.D. Auret, W. Mtangi, M. Diale, J.M. Nel, Thermal annealing behaviour of platinum, nickel and titanium Schottky barrier diodes on n-Ge (100), Journal of Alloys and Compounds 492 (2010) 649–655, doi:10.1016/j.jallcom.2009.11.202

[23] A. Chawanda, S.M.M. Coelho, F.D. Auret, W. Mtangi, C. Nyamhere, J.M. Nel, M. Diale, Effect of thermal treatment on the characteristics of iridium Schottky barrier diodes on n-Ge (100), Journal of Alloys and Compounds 513 (2012) 44–49,
https://doi.org/10.1016/j.jallcom.2011.09.053

[24] A. Ohta, M. Matsui, H. Murakami, S. Higashi, S. Miyazaki, Control of Schottky barrier height at Al/p-Ge junctions by ultrathin layer insertion, ECS Transactions, 50 (9) 449-457 (2012), https://iopscience.iop.org/article/10.1149/05009.0449ecst/meta

[25] P.S.Y. Lim, D.Z. Chi, X.C. Wang, Y.-C. Yeo, Fermi-level depinning at the metal germanium interface by the formation of epitaxial nickel digermanide $NiGe_2$ using pulsed laser anneal, Appl. Phys. Lett. 101, 172103 (2012) https://doi.org/10.1063/1.4762003

[26] A. Firrincieli, K. Martens, E. Simoen C. Claeys, J. A. Kittl, Study of the impact of doping concentration and Schottky barrier height on ohmic contacts to n-type germanium, Microelectronic Engineering 106 (2013) 129-131, https://doi.org/10.1016/j.mee.2012.12.020

[27] K. Gallacher, P. Velha, D. J. Paul, I. MacLaren, M. Myronov, D. R. Leadley, Ohmic contacts to n-type germanium with low specific contact resistivity, Appl. Phys. Lett. 100 (2012) 022113, https://doi.org/10.1063/1.3676667

[28] R. Li, H.B. Yao, S.J. Lee, D.Z. Chi, M.B. Yu, G.Q. Lo, D.L. Kwong, Metal-germanide Schottky Source/Drain transistor on Germanium substrate for future CMOS technology, Thin Solid Films 504 (2006) 28-31, https://doi.org/10.1016/j.tsf.2005.09.033

[29] V. Marrello, T.A. McMath, J.W. Mayer and I.L. Fowler, High purity germanium gamma-ray spectrometers with regrowth p+ contacts, Nuclear Instruments and Methods 108 (1973) 93-97, https://doi.org/10.1016/0029-554X(73)90640-X

[30] A. Habanyama, Interface Control Processes for Ni/Ge and Pd/Ge Schottky and Ohmic Contact Fabrication: Part One, Advanced material and device applications with germanium, ISBN: 978-1-83881-724-4, http://dx.doi.org/10.5772/intechopen.78692

[31] İ. Taşçıoğlu, U. Aydemir, Ş. Altındal, B. Kınacı, S. Özçelik, Analysis of the forward and reverse bias *I-V* characteristics on Au/PVA:Zn/n-Si Schottky barrier diodes in the wide temperature range, J. Appl. Phys. 109 (2011) 054502, https://doi.org/10.1063/1.3552599

[32] I. Jyothi, H.-D. Yang, K.-H. Shim, V. Janardhanam, S.-M. Kang, H. Hong, C.-J. Choi, Temperature dependency of Schottky barrier parameters of Ti Schottky contacts to Si-on-insulator, Materials Transactions 54 (9) (2013) 1655-1660,
 https://doi.org/10.2320/matertrans.M2013015

[33] P. Reddy I. Bryan, Z. Bryan, J. Tweedie, S. Washiyama, R. Kirste, S. Mita, R. Collazo, Z. Sitar, Charge neutrality levels, barrier heights, and band offsets at polar AlGaN, Appl. Phys. Lett. 107 (2015) 091603, https://doi.org/10.1063/1.4930026

[34] G.-P. Ru, R.L.V. Meirhaeghe, S. Forment, Y-L. Jiang, X.-P. Qu, S. Zhu, B.-Z. Li, Voltage dependence of effective barrier height reduction in inhomogeneous Schottky diodes, Solid-State Electronics 49 (2005) 606–611, doi:10.1016/j.sse.2004.12.005

[35] P. J. King, E. Arac, S. Ganti, K. S. K. Kwa, N. Ponon, A. G. O'Neill, Improving metal/semiconductor conductivity using AlOx interlayers on n-type and p-type Si, Appl. Phys. Lett. 105 (2014) 052101, https://doi.org/10.1063/1.4892003

[36] M. Ghosh, S. Pitale, S.G. Singh, H. Manasawala, V. Karki, M. Singh, K. Singh, G.D. Patra, S. Sen, Fabrication of *p+* contact by thermally induced solid state regrowth of *Al* on *p-type Ge* crystal, Materials Science in Semiconductor Processing 121 (2021) 105350, https://doi.org/10.1016/j.mssp.2020.105350